\documentclass[a4paper,fleqn,usenatbib]{mnras}

\usepackage{newtxtext,newtxmath}
\usepackage[T1]{fontenc}
\usepackage{ae,aecompl}
\usepackage{graphicx}	
\usepackage{amsmath}	
\usepackage{amssymb}	
\usepackage{color}

\title[CO-poor/C{\sc i}-rich $\rm H_2$ gas in the Universe]{New places and phases of CO-poor/C{\sc i}-rich molecular gas in the Universe}

\author[Papadopoulos et al.]{
Padelis P. Papadopoulos,$^{1,2,3}$\thanks{E-mail: padelis@auth.gr}
Thomas~G. Bisbas,$^{4,5}$
and Zhiyu Zhang$^{6,7}$
\\
$^{1}$Department of Physics, Section of Astrophysics, Astronomy and Mechanics,Aristotle University of Thessaloniki,\\ Thessaloniki, GR-54124, Greece\\
$^{2}$Research Center for Astronomy, Academy of Athens
 Soranou Efesiou 4, GR-11527, Athens, Greece\\
$^{3}$School of Physics and Astronomy, Cardiff University,
 Queen's Buildings, The Parade, Cardiff, CF24 3AA, UK\\
$^{4}$Department of Astronomy, University of Virginia, Charlottesville, VA 22904, USA\\
$^{5}$Max-Planck-Institut f\"ur Extraterrestrische Physik, Giessenbachstrasse 1, D-85748 Garching, Germany\\
$^{6}$Institute for Astronomy, University of Edinburgh, Royal Observatory, Edinburgh, EH9 3HJ, UK\\
$^{7}$European Southern Observatory, Headquarters, Karl-Schwarzschild-Strasse 2, D-85748, Garching bei M\"unchen, Germany}

\date{Accepted XXX. Received YYY; in original form ZZZ}

\pubyear{2015}

\begin{document}
\label{firstpage}
\pagerange{\pageref{firstpage}--\pageref{lastpage}}
\maketitle

\begin{abstract}
In this  work we extend  the work on  the recently discovered  role of
Cosmic Rays  (CRs) in  regulating the  average CO/$\rm  H_2$ abundance
ratio  in molecular  clouds (and  thus  their CO  line visibility)  in
starburst  galaxies, and  find that  it  can lead  to a  CO-poor/C{\sc
  i}-rich $\rm H_2  $ gas phase even in environments  with Galactic or
in  only  modestly  enhanced   CR  backgrounds  expected  in  ordinary
star-forming galaxies. Furthermore, the same CR-driven astro-chemistry
raises  the  possibility of  a  widespread phase  transition  of
  molecular gas towards a CO-poor/C{\sc i}-rich phase in: a) molecular
  gas  outflows found  in star-forming  galaxies, b)  active galactic
nuclei (AGNs),  and c)  near synchrotron-emitting  radio jets  and the
radio-loud  cores  of  powerful  radio galaxies.   For  main  sequence
galaxies we find that CRs can  render some of their molecular gas mass
CO-invisible, compounding  the effects  of low  metallicities. Imaging
the two  fine structure  lines of atomic  carbon with  resolution high
enough to search beyond the  C{\sc i}/CO-bright line regions associated
with  central  starbursts  can  reveal such  a  CO-poor/C{\sc  i}-rich
molecular gas  phase, provided that relative brightness  sensitivity levels
of $T_b$(C{\sc i} $1-0$)/$T_b$(CO $J=1-0$)$\sim $0.15 are reached. The
capability to search  for such gas in  the Galaxy is now  at hand with
the new  high-frequency survey  telescope HEAT deployed  in Antarctica
and future ones  to be deployed in  Dome A.  ALMA can  search for such
gas in star-forming spiral disks,  galactic molecular gas outflows and
the CR-intense  galactic and  circumgalactic gas-rich  environments of
radio-loud objects.
\end{abstract}




\section{Introduction} \label{sec:intro}

The utility of CO and its low-$J$ rotational transitions as effective tracers
of $\rm H_2$ gas mass is now well established observationally
\citep[e.g.][]{Youn91, Solo92, Solo97} and theoretically
\citep[e.g.][]{Dick86,Malo88,Brya96}, provided: a) the so-called $X_{\rm
CO}(=M({\rm H}_2)/L_{\rm CO})$ factor is used only for $M({\rm H}_2)\geq
10^{5}\,M_{\odot}$ (so that its statistical notion remains valid) and b) the
$\rm [CO]/[H_2]$ abundance ratio does not fall much below its average Galactic
value of $\sim 10^{-4}$. The latter can happen in low-metallicity interstellar
medium (ISM) with strong ambient far-ultraviolet (FUV) radiation fields
\citep[][]{Bola99,Pak98}, such as those expected in metal-poor dwarf galaxies,
leaving large amounts of $\rm H_2$ gas as CO-invisible \citep{Madd97,Shi17}. In
the Milky Way and other ordinary spirals, this is expected also for molecular
gas at large galactocentric distances where metallicity falls to $\sim 0.2 \rm
Z_{\odot}$ and much of the $\rm H_2$ gas can be rendered (CO line)-invisible by
the FUV-induced destruction of CO \citep{Papa02,Wolf10}.

It was only recently that Cosmic Rays (CRs) have been identified as a potentially
more effective CO-destruction agent in molecular clouds compared to FUV
photons \citep{Bisb15,Bisb17,Bial15}. Unlike the latter whose propagation (and
thus CO-dissociation capability) is blunted by the strong dust absorption of
FUV light taking place in the dust-rich H{\sc i} phase and in outer $\rm H_2$
cloud envelopes, CR-induced chemistry destroys CO volumetrically throughout a
molecular cloud irrespective of dust column. Other observational signatures
  of  CR-controlled versus FUV-controlled chemistry of H$_2$ clouds in galaxies,
  (even when CO remains abundant) have been discussed thoroughly in the literature
  \citep[e.g.][]{Papa10,Meij11}, and will not concern us here.
We should, nevertheless, mention that it is CRs that make the
 C{\sc i} distribution concomitant with that of CO in H$_2$ gas clouds, rather
existing only in a thin transition layer between C{\sc ii}-rich outer and CO-rich
inner H$_2$ cloud regions \citep[as the traditional PDR view would have it, e.g.][]{Holl99}. 
This makes  C{\sc i} lines equally good and  more straightforward  H$_2$ gas mass
tracers as low-J CO lines, even under conditions where CO remains abundant
in H$_2$ gas clouds.

A last theoretical effort to retain the PDR picture against the failure
of its basic prediction of a C{\sc ii}/C{\sc i}/CO stratification of species on
the surface of FUV-illuminated H$_2$ clouds was made by introducing
 density inhomogeneities on the classic PDR picture \citep{Meix93,Spaa97}.
 There, the C{\sc ii}/C{\sc i}/CO species stratification remained but only on small H$_2$ clumps
while C{\sc i} appeared spatially extended deeper into FUV-illuminated CO-rich inhomogeneous
H$_2$ clouds. However, while CO-rich H$_2$ clumps  of low filling factor, each with a
C{\sc i} ``coating", could reproduce  the astonishing spatial correspondence  between C{\sc i} and
$^{12}$CO,  $^{13}$CO line  emission observed  in GMCs  across a  wide range  of
conditions, they could not account for the tight {\it intensity correlation}
between $^{12}$CO,  $^{13}$CO (1-0), (2-1) and  C{\sc i} 1-0 line intensities,  unless one
postulates also a very standard H$_2$ clump making up all H$_2$ clouds, with
characteristics  that  remain  invariant  across  the  wide  range  of
ISM conditions \citep[see][for details]{Papa04}. CRs are  the simplest and  most likely  culprits in
creating a  volumetric rather than  surface-like C{\sc i} distribution  in H$_2$
clouds,  and   therein  lies  the  most   notable  difference  between
FUV-driven chemistry and  gas thermal state, and a  CR-driven one.

Another key difference between the two mechanisms besides their spatially distinct 
ways to destroy CO is that FUV-induced CO destruction leaves behind C{\sc i} and 
C{\sc ii}, while CR-induced destruction yields mostly C{\sc i}, provided that $T_{\rm
kin}\lesssim50\,{\rm K}$, otherwise CO abundance increases via the OH channel
\citep{Bisb17}. This makes CR-induced CO-poor $\rm H_2$ gas more accessible to
observations via the two fine structures of atomic carbon at rest frequencies
$\nu_{\rm rest}^{\rm CI}(1-0) \sim $492\,GHz $^3{\rm P}_1-^3{\rm P}_0$
(hereafter 1-0) and $\nu_{\rm rest}^{\rm CI}(2-1) \sim $809\,GHz $^3{\rm
P}_2-^3{\rm P}_1$ (hereafter 2-1) than FUV-irradiated clouds. These [C{\sc i}]
lines can be observed over a large redshift range, starting from $z\sim $0 to
$z\sim $5 for [C{\sc i}] 1-0, and from $z\sim $0 to $z\sim $8 for [C{\sc i}]
2-1, using ground-based telescopes, such as ALMA on the Atacama Desert Plateau,
while the [C{\sc ii}] fine structure line, even if typically much brighter than
the [C{\sc i}] lines for warm gas, will be faint for cold $\rm H_2$ gas
($T_k\sim (15-20)\,{\rm K}$, while $\rm E_{ul}($C{\sc ii}$\rm )/k_B\sim $92\,K)
away from star-forming (SF) sites. Moreover C{\sc ii} has a rest frequency of
$\sim $1900\,GHz, making it accessible to ground-based observations only once
$\rm z\ga 2$, still extremely challenging until $\rm z\ga 4$. This leaves most
of the star-formation history of the Universe (and its gas-fueling) outside the
reach of [C{\sc ii}].

In this paper we study the effects of a CR-regulated [CO/C{\sc i}]
average abundance in low-density molecular gas in the Galaxy, the
outer regions of local spirals, and distant main sequence (MS)
galaxies. We conclude this work by examining the possibility of
CO-poor/C{\sc i}-rich molecular gas in the CR-intense environments of
molecular gas outflows from starbursts, and the environments of radio-loud
objects.

\section{Low-density molecular gas in the Universe: the effects of CRs}
\label{sec:2}

The CR effects on the relatively low density molecular gas ($n({\rm H_2})
\sim$50-500\,${\rm cm}^{-3}$) have not been studied in detail, but early hints
that CO can be effectively destroyed in such gas {\it even at Galactic levels
of CR energy densities} exist \citep{Bisb15}.  A low density 
molecular gas phase can be found in a variety of places in the Universe, the nearest 
ones being the envelopes of ordinary GMCs in the Galaxy. Should their CO-marker 
molecule be wiped out by CRs, it would leave the corresponding H$_2$ gas mass 
CO-invisible, yielding a systematic underestimate of H$_2$ gas mass even in places 
where CO was considered an effective $\rm H_2$ tracer. This is of particular 
importance since a typical log-normal distribution of $M$(H$_2$)-$n$($\rm H_2$) 
expected in turbulent GMCs would place most of their total mass at densities of 
$n({\rm H}_2)<500\,{\rm cm}^{-3}$ \citep{Pado02}. Moreover, CRs can act
  on the chemistry of H$_2$ gas as fast as the photon-driven processes driven by
  FUV radiation fields.

\subsection{The Milky Way}
\label{ssec:MW}

Studies of the H{\sc i}$\rightarrow $H$_2$ phase transition in the metallicity
and radiation environment of the Milky Way showed that it can commence from
densities as low as $n({\rm H_2}) \sim$ 5--20\,${\rm cm}^{-3}$, depending on
the $\rm H_2$ formation rate on grains and ambient dust shielding
\citep{Papa02}, while for density enhancements reaching above $n({\rm H_2})
\sim$50\,${\rm cm}^{-3}$ this transition is complete
\citep{Jura75a,vDis86,Jura75b,Shay87,Ande93,Shul00,Offn13,Bial17}. However CO
(and HCN) multi-$J$ observations of Molecular Clouds (MCs) in the Galaxy,
typically yield densities of $n({\rm H_2}) \sim$ 500--$10^4$\,${\rm cm}^{-3}$
\citep{Saka97,Heye15,Bial16}. Thus, there is a significant range of
gas densities $n({\rm H_2}) \sim$ 50--500\,${\rm cm}^{-3}$ where Cold Neutral
Medium (CNM) gas can be molecular but perhaps  not  (CO-line)-bright, reminiscent
of the translucent clouds \citep{vDis86}.

\begin{figure*}
\centering
\includegraphics[width=\textwidth]{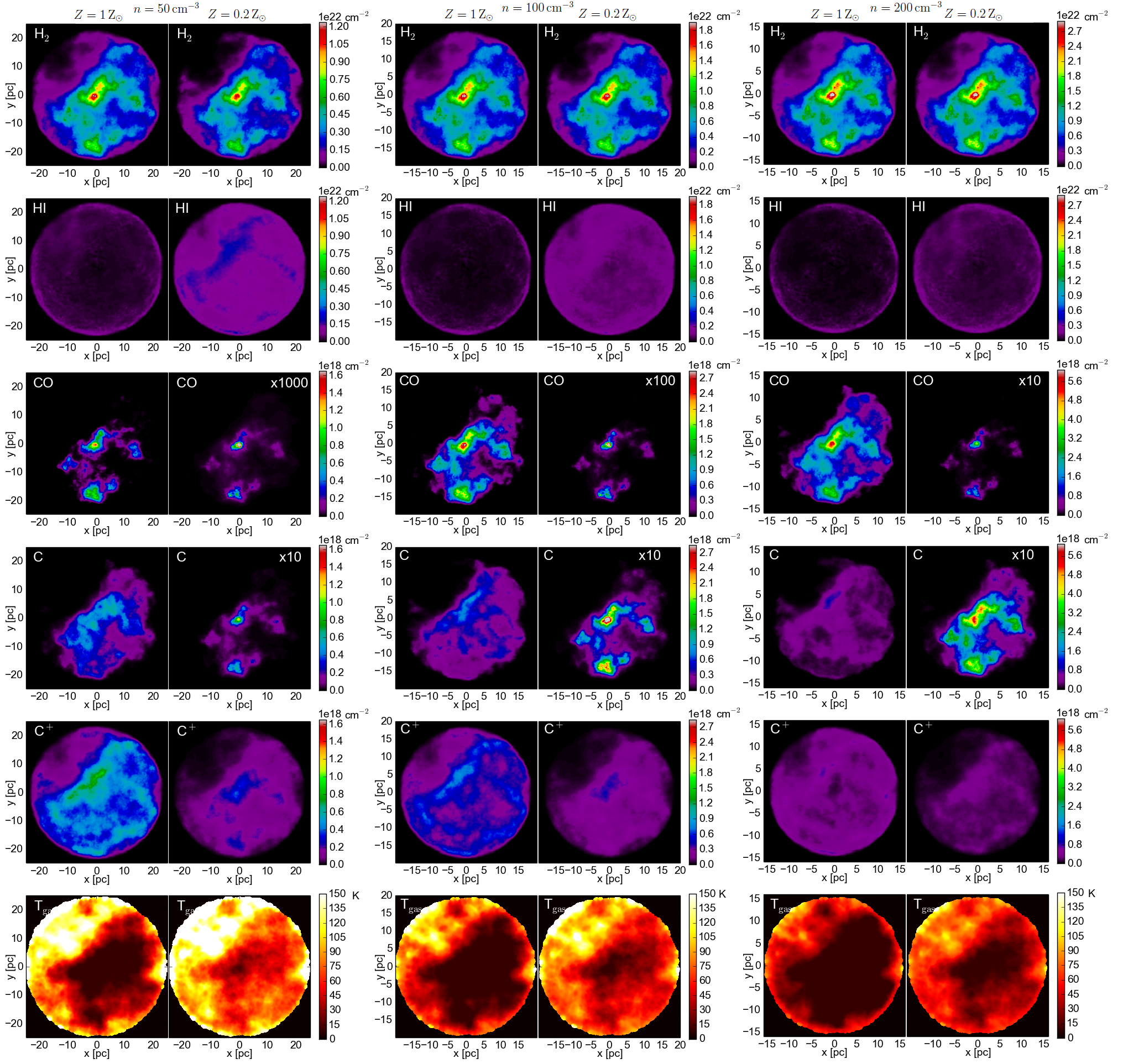}
\caption{Column density ($N$) plots of H$_2$ (top row), H{\sc i} (second row),
CO (third row), C{\sc i} (fourth row), C{\sc ii} (fifth row) in different
volume H$_{2}$ densities and metallicities. The color bar has units of
cm$^{-2}$ and the axes have units of pc. The bottom row shows cross sections of
the gas temperature at the $z=0\,{\rm pc}$ plane. The color bar there has units
of K. The first two columns correspond to the GMC with $\langle
n\rangle\sim50\,{\rm cm}^{-3}$, the middle two columns to the GMC with $\langle
n\rangle\sim100\,{\rm cm}^{-3}$ and the right two columns to the GMC with
$\langle n\rangle\sim200\,{\rm cm}^{-3}$. In each pair of columns, the left one
corresponds to solar metallicity ($Z=1Z_{\odot}$) and the right one to
sub-solar metallicity ($Z=0.2Z_{\odot}$). In all cases, we consider a Galactic
CR energy density ($\zeta'=1$) and ${\rm G}_{\circ}=1$, normalized according to
\citet{Drai78}. It can be seen that $N({\rm H}_2)$ remains unchanged with
decreasing $Z$ and that $N$(C{\sc i}) and particularly $N$(C{\sc ii}), do not
decrease as dramatically as $N$(CO) (note the scaling factors in the 
$0.2\,Z_{\odot}$ panels). As expected, the gas temperature slightly
increases with decreasing $Z$, since the absence of metals reduces the overall
cooling in the GMCs.} 
\label{fig:cds}
\end{figure*}

Figure~\ref{fig:cds} shows the column density maps of H{\sc i}, $\rm H_2$ CO,
C{\sc i}, C{\sc ii}, and $T_{\rm kin}$ distributions within inhomogeneous
low-density gas clouds using the fractal rendering and thermo-chemical
calculations presented by \citet{Bisb17} and using the {\sc 3d-pdr}
code\footnote{https://uclchem.github.io/} \citep{Bisb12}, but now subjected
only to Galactic levels of interstellar radiation field and CR energy density
\citep[i.e. $\zeta'=1$, where $\zeta'=\zeta_{\rm CR}/10^{-17}\,{\rm s}^{-1}$
and $\rm G_{\circ}=1$ normalized according to][]{Drai78}. 
The GMCs were constructed using the method described in \citet{Walc15} for a 
fractal dimension of ${\cal D}=2.4$ and assuming a mass of
$M=7\times10^4\,{\rm M}_{\odot}$ but with different radial extent corresponding
to three average number densities i.e. $\langle n\rangle\sim50,\,100,\,200\,{\rm 
cm}^{-3}$. We find that the maximum visual extinction, $A_V$, along the line-of-sight 
of these clouds is $\sim14,\,22,\,40\,{\rm mag}$, respectively.
These GMCs have much smaller average number densities than the GMC
studied in \citet{Bisb17} ($\langle n\rangle\sim760\,{\rm cm}^{-3}$).
Furthermore, our computations are made for metallicities of both $Z=Z_{\odot}$ and
$Z=0.2Z_{\odot }$ representing the ISM for inner and the outer parts of the Milky Way. 

From the maps in Figure~\ref{fig:cds} it can be readily seen that for $\langle
n\rangle\sim(50-100)\,{\rm cm}^{-3}$ {\it the $\rm H_2$ gas can be rendered very
CO-poor, even at Galactic levels of CR energy density}, while for low
metallicities this remains so up to $\langle n\rangle\sim200\,{\rm cm}^{-3}$
(the highest average density in our computations). Atomic Carbon on the other
hand remains abundant throughout the H$_2$-rich parts of the cloud, except for
the low-metallicity gas where its abundance drops, but nevertheless remains
 generally higher than that of CO. 
The cloud mass fractions for which the [CO]/[H$_2$] abundance drops
below $10^{-5}$ (i.e. 10 times below the average [CO]/[H$_2$]
abundance in the Galaxy, making hard to use CO lines as $\rm H_2$ gas
mass tracers) are: 80\%, 55\%, 30\% for $\langle
n\rangle\sim50,\,100,\,200\,{\rm cm}^{-3}$, respectively, for
$Z=1Z_{\odot}$, K$_{\rm vir}=1$\footnote{With $\rm
 K_{vir}=(dV/dR)/(dV/dR)_{vir}$ a value of $\rm K_{vir}=1$ signifies
 self-gravitating clouds, see \citep{Papa14} for details} and
$\zeta'=1$. For the metal-poor case, the CO-poor cloud mass fraction
becomes $\gtrsim 98\%$.

Nevertheless, unlike in our past higher-density cloud models where
most of C{\sc ii} recombines into C{\sc i} \citep{Bisb17}, C{\sc ii}
now remains abundant for much of the mass of our low-density clouds
(Figure~\ref{fig:cds}). Furthermore, classical photolectric FUV
heating, and the much lower average cooling of low density gas
($\Lambda _{\rm line}\propto n^2$) allows the gas to maintain higher
temperatures $T_{\rm kin}\sim(40-100)\,{\rm K}$ for $\langle
n\rangle\sim(50-100)\,{\rm cm}^{-3}$, where the [C{\sc ii}] fine
structure line is expected to be luminous.

\begin{figure*}
\centering
\includegraphics[width=0.8\textwidth]{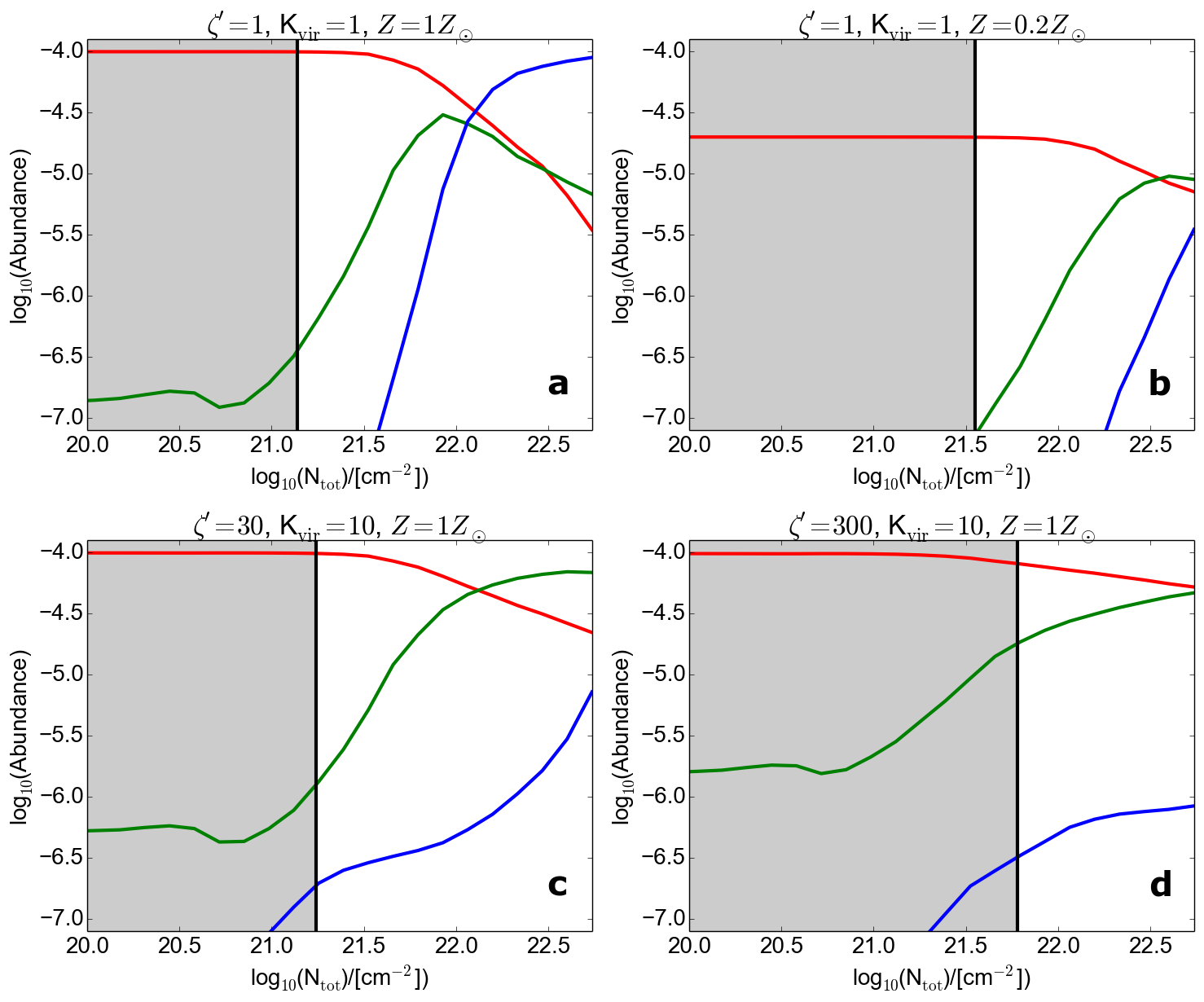}
\caption{The abundances of C{\sc ii} (red line), C{\sc i} (green line) and CO (blue line) 
versus the total H-nucleus column density. Each line corresponds to the average value obtained
from all three GMCs. Panels a and b correspond to the bound (K$_{\rm vir}=1$)
GMC embedded in Galactic CR energy density ($\zeta'=1$). We consider solar
metallicity ($Z=1Z_{\odot}$; panel a) and sub-solar metallicity
($Z=0.2Z_{\odot}$; panel b). Panels c and d correspond to the unbound GMC
(K$_{\rm vir}=10$) for solar metallicity. In panel c the CR
ionization rate is taken to be 30 times the Galactic ($\zeta'=30$) and in panel
d 300 times ($\zeta'=300$). The vertical line shows the H{\sc i}-to-H$_2$ transition 
(shaded area is atomic). We find that CO dominates as the main repository of carbon only for
Galactic conditions (panel a) even as C remains abundant enough ($\rm [CI/CO]\sim 0.1-0.5$ for
the bulk of $\rm H_2$ mass) as to remain a capable molecular gas mass tracer along with low-J CO lines.}
\label{fig:cdsall}
\end{figure*}

Figure~\ref{fig:cdsall} shows how the carbon cycle abundances change
as a function of the total H column density for the four different ISM
models. Here, we plot the average value of all three different clouds simultaneously.
It can be seen that for the Galactic conditions (panel a), the molecular gas is
CO-dominated for high column densities, as expected, while C remains
abundant enough ($\rm [CI/CO]\sim 0.1-0.3$) as to continue serving as
a capable $\rm H_2$ gas tracer along with CO \citep[e.g.][]{Papa04}.
However, the molecular gas phase switches to a C{\sc i}-dominated one in the
lower-metallicity case (panel b) and for  $Z=1\,Z_{\odot}$ and $\zeta'=30$ case (panel c).
For even higher $\zeta'$, panel (d) shows a gas phase that becomes C{\sc ii}-dominated
even at higher column densities  (i.e. $\sim6\times10^{21}\,{\rm cm}^{-2}$) more typical
for inner regions of molecular clouds.

Thus it may well be that, besides FUV photons, CRs also contribute in
the making of (C{\sc ii}-line)-bright gas envelopes of (CO
line)-invisible gas found around (CO line)-marked GMCs in the Galaxy
\citep[e.g.,][]{Pine13,Lang14}. {\it Then these envelopes can be
 bright in both C{\sc ii} and C{\sc i} lines,} a possibility that
should be investigated by sensitively imaging the latter in the C{\sc
 ii}-bright envelope regions of GMCs in the metal-rich inner parts of the
Galaxy. The conditions for a widespread phase transition of
$\rm H_2$ gas from a CO-rich to a CO-poor/CI-rich phase can exist also
  for outer Galactic regions. Indeed as the star formation rate density
becomes lower in the outer Galaxy, along with the intervening
interstellar absorption, they reduce the average FUV radiation field
at large Galactocentric distances.  CRs however can  stream further
out in the disk, and keep C{\sc i} abundant in  low-density molecular clouds.

There is already evidence for  CO-poor gas in the Galaxy at large
  galactocentric  radii from  studies  of  otherwise CO-bright  clouds
  selected in the Goddard-Columbia  $^{12}$CO survey.  In these places
  an $\rm  X_{CO}$ factor systematically  larger by a factor  of $\sim
  2-3$ with respect to its standard value at the inner Galaxy is found
  \citep{Sodr91}. Other work using also CO-marked clouds finds
  that the diffuse  H$_2$ clouds contain increasingly more  $\rm H_2 $
  mass at larger Galactocentric radii \citep{Roma16}. These
  are  the   type  of   clouds  that   could  contain   a  significant
  CI-rich/CO-poor phase  according to our current  study. Perhaps more
  significantly, the only  search for the so so-called  CO-dark gas in
  our Galaxy which is not using  CO-selected $\rm H_2$ clouds but dust
  extinction maps  instead indicates  that up  to 55\%  of the  gas at
  large Galactocentric radii may be  CO-dark \citep{Chen15}.  The
  same study finds  that in certain regions the  inferred CO-dark $\rm
  H_2$ gas mass can reach up to four times the CO-luminous one.

  We  must  note  here   that  all  our  current  arguments  about
  CO-poor/CI-rich  gas, as  well  as  those that  follow  in the  next
  sections, are  phase-transition type of  arguments. By this  we mean
  that we are simply investigating  the ISM conditions controlling the
  phase transition from a typically CO-rich to a CO-poor (and CI-rich)
  gas  phase (with  emphasis  on the  role of  CRs,  gas density,  and
  metallicity), and the  places where the conditions for  such a phase
  transition can  be fullfiled.  In order  to find how much  $\rm H_2$
  mass   is  actually   in  such   a  gas   phase,  observations   are
  indispensible.  To this  purpose we conclude this  section by urging
  sensitive imaging of the two C{\sc i} lines for the molecular gas in
  the Galaxy, and especially of its CO-poor yet CII-luminous phase.

\subsubsection{The promise of ground-based high-frequency single dish telescope surveys}

The High Elevation Antarctic Terahertz Telescope (HEAT), with a
diameter of 60 cm, is a new high-frequency telescope now operating at
Ridge A of the Antarctic plateau, with capabilities to observe both
[C{\sc i}]
lines\footnote{http://soral.as.arizona.edu/HEAT/instrument/}. Its wide
beam of $\sim 4.1^{'}$ at 492\,GHz ([C{\sc i}] 1-0) and $2.5^{'}$ at
809\,GHz ([C{\sc i}] 2-1) makes it appropriate for searching for
low-brightness temperature (T$_{\rm b}$) C{\sc i}-rich molecular
clouds in the Galaxy, where 
$T_b$([C{\sc i}] 1-0)/$T_b$(CO 1-0) (hereafter $T_b$(C{\sc
  i})/$T_b$(CO)) drops to $\sim 0.15$ in SF-quiescent regions
\citep[][and references  therein]{Papa04}. An inventory of molecular gas in the Galaxy
obtained using both [C{\sc i}] lines and its comparison to the CO-rich
and C{\sc ii}-rich gas can reveal whether a low-density C{\sc
 i}-rich/CO-poor gas phase exists in the Milky Way, its spatial
distribution, and temperature.

\begin{figure*}
\centering
\includegraphics[width=0.8\textwidth]{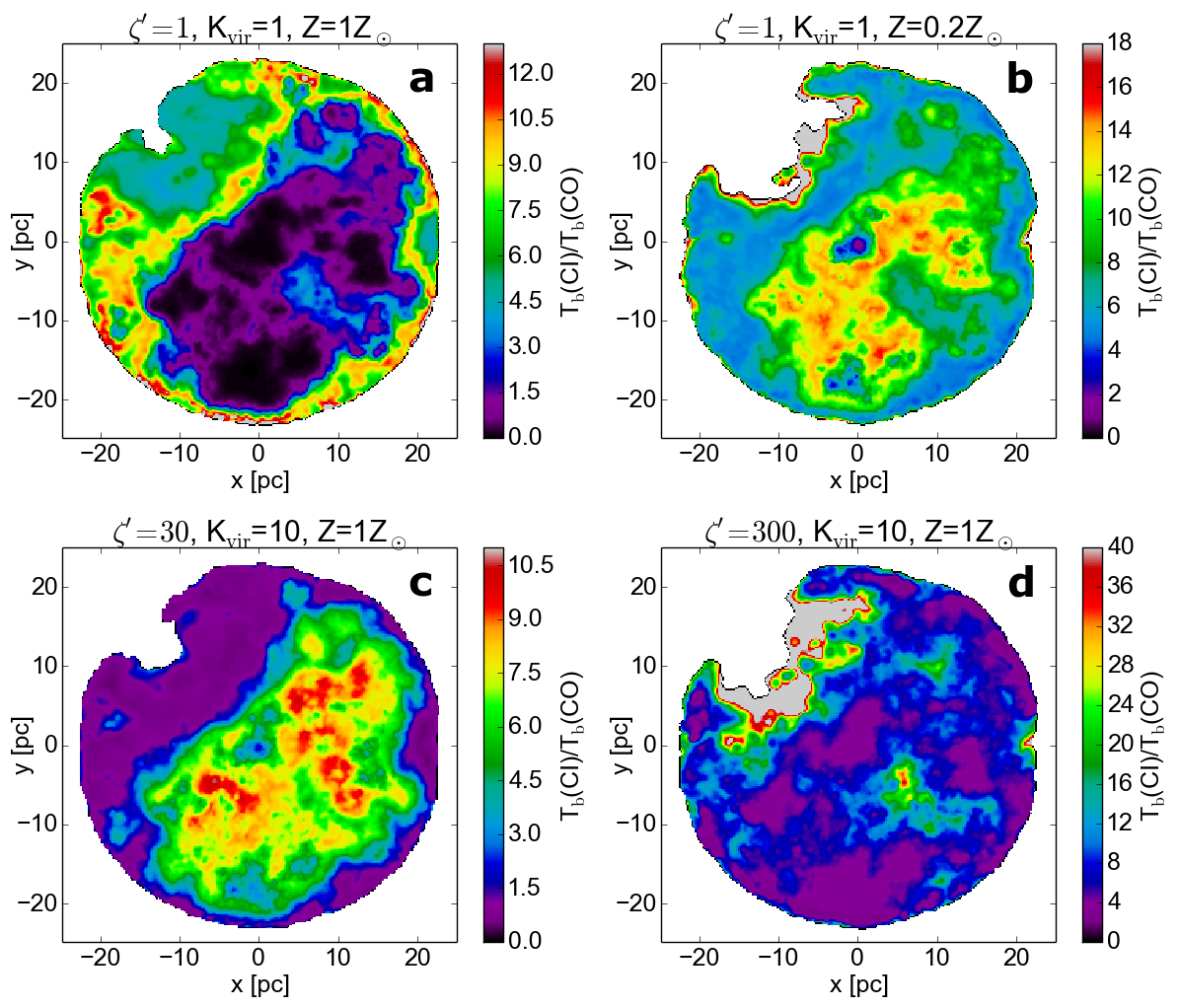}
\caption{Maps of T$_{\rm b}$(C{\sc i})/T$_{\rm b}$(CO) (for the lowest $1-0$
transition) brightness ratio of the GMC with $\langle n\rangle\sim50\,{\rm
cm}^{-3}$ for different conditions. Panels a-d follow the ISM and GMC conditions described
in Fig.~\ref{fig:cdsall}. In all cases we considered an isotropic
interstellar radiation field with strength $G_{\circ}=1$, normalized according
to \citet{Drai78}. The corresponding N(H$_2$) distribution of panels a and b can be
seen in the top row, left pair of columns of Fig.~\ref{fig:cds}.
The corresponding N(H$_2$) distribution of panels c and d can be
seen in the top row, left pair of columns of Fig.~\ref{fig:highzcr}.
We find that lower H$_2$ column densities
($N$(H$_2$)$\lesssim10^{22}\,{\rm cm}^{-2}$, see Fig.~\ref{fig:panels}) are in
general brighter in CO in the Milky Way-type of ISM conditions. In all other
cases, we predict that these column densities will be brighter in [C{\sc i}]
(1-0) than in CO (1-0). Note also that in panels c and d, much lower column
densities may become brighter in CO (1-0); this is because of the formation of
CO via the OH channel as explained in detail in \citet{Bisb17}, which locally
increases the abundance of CO and hence its emission.} 
\label{fig:heat}
\end{figure*}

Other single dish submm telescopes from excellent sites at the Atacama
Plateau in North Chile (APEX, NANTEN-2, ASTE) can be used to
sensitively map C{\sc i} (1-0) and/or (2-1) in a more targeted fashion
e.g. along a continuous strip starting from the inner parts of a
given Galactic molecular cloud and continuing well beyond its
(CO/CI-line)-luminous regions, searching for C{\sc i}-rich/CO-poor
envelopes. The upcoming 5-m Dome A Terahertz Explorer
\citep[DATE5][]{Yang13} in Dome-A of the Antarctic plateau will be
able to perform a systematic survey across the Galactic plane with both
good angular resolution and sensitivity, given its excellent condition
for high-frequency observations\citep{Shi16}. In
Figure~\ref{fig:heat}, we show $T_b$(C{\sc i})/$T_b$(CO) ratio maps
for our low-density ($\langle n\rangle$$\sim$50\,${\rm cm}^{-3}$)
inhomogeneous cloud model. These could be used to guide such an
observational campaign in the Galaxy indicating the necessary
CI/CO line relative brightness sensitivity levels.

\subsection{Molecular gas outflows from galaxies}
\label{ssec:outflows}

The discovery of strong H$_2$ outflows from galaxies, induced by AGN
and/or starburst activity \citep{Feru10,Cico12,Cico14,Dasy12}, shows
that large amounts of molecular gas can be expelled from galaxies. It
followed much earlier discoveries of huge cm-emitting synchrotron
haloes around starburst galaxies \citep[e.g.][]{Seaq91,Colb96},
indicating CRs outflows swept out from the star-formation (SF) galaxy
at bulk speeds $\ga 1000$\,km\,s$^{-1}$. This could have significant
effects since powerful molecular outflows are often discovered in
those same starbursts where extended synchrotron haloes are found
\citep[e.g.][]{Sala13}. Thus, {\it the CO-destroying CRs could be
 swept along the molecular gas outflow itself.} Moreover, by
exerting large pressures on the molecular gas of starburst galaxies
(even small gas ionization fractions can allow CR-gas coupling and
momentum transfer from CRs to gas), CRs could even be driving these
fast molecular outflows \citep{Hana13,Giri16}.

Low-density and gravitationally unbound molecular gas is to be
expected in such galactic outflows, a result of Kelvin-Helmholtz
instabilities and shear acting on the envelopes of denser clouds in
the outflow. Such a gas phase could carry a significant mass fraction
of the outflow, while remaining CO-invisible, because of large scale CO destruction
induced by the CRs carried within the same outflow. Large masses of low-density molecular
gas could be present in galactic molecular gas outflows given the 
trend of progressively larger amounts of molecular gas mass discovered
in them, the lower the critical density of the line tracer
used to reveal them is \citep{Cico12}. A
CR-irradiation of the outflowing molecular gas by the relativistic
plasma carried along with it thus points to the possibility of
CO-poor/C{\sc i}-rich molecular gas  in powerful galactic
outflows.

Using the magnetic field value found for the outflow in M\,82 of $\rm \langle
B\rangle=25\,\mu$G \citep{Adeb13}, and the equipartition
assumption between magnetic field and CR energy densities, yields a CR energy
density boost expected in such an outflow of $U_{\rm CR}=(\langle {\rm
B}\rangle/\langle {\rm B}_{\rm Gal}\rangle)^2\times U_{\rm CR,Gal}\sim 17\times
U_{\rm CR,Gal}$ (for $\langle \rm B_{Gal}\rangle\sim 6\,\mu$G). Stronger
magnetic fields of $\rm \langle B\rangle\sim(35-40)\mu$m have been found in
extended synchrotron haloes around galaxies \citep{Lain08}, corresponding to
$U_{\rm CR}\sim (34-44)\times U_{\rm CR,Gal}$ for any concomitant $\rm H_2$ gas
phase outflowing along with the CRs. For our computations, we adopt $U_{\rm
CR}=30\times U_{\rm CR,Gal}$ and $\rm K_{vir}=10$ corresponding to strongly
unbound gas states expected in such galactic outflows.

\begin{figure*}
\centering
\includegraphics[width=\textwidth]{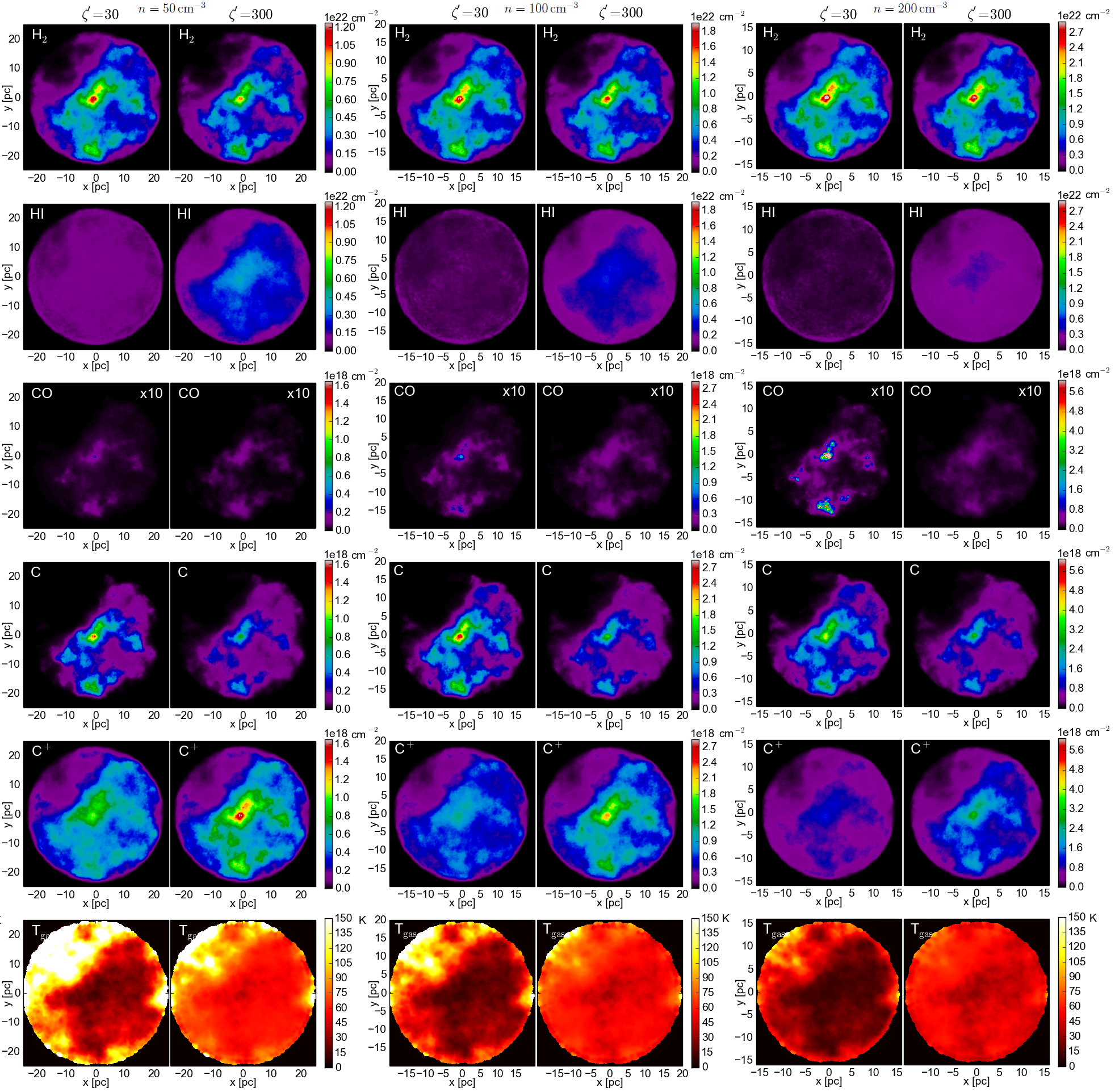}
\caption{As in Fig.~\ref{fig:cds}, but here the left column of each pair corresponds to an elevated CR energy density of $\zeta'=30$ times the Galactic, and the right column to $\zeta'=300$ times the Galactic. We consider $Z=1Z_{\odot}$ and $G_{\circ}=1$ everywhere. It can be seen that $N$(H$_2$) remains remarkably unchanged at all times. $N$(CO) is always smaller than $N$(C{\sc i}) and $N$(C{\sc ii}), which is in agreement with the findings of \citet{Bisb15,Bisb17}. The gas temperature at the interior of the cloud increases with increasing $\zeta'$ and in the particular case of $\zeta'=300$, it is approximately uniform ($T_{\rm gas}\sim50-70\,{\rm K}$) regardless of the local density, $n$. The conditions presented here are expected to be found in radio-galaxies.}
\label{fig:highzcr}
\end{figure*}

Our results are shown in Figure~\ref{fig:highzcr}, from where it can
be seen that CO is destroyed very effectively over the whole density
range we consider in this work, while C{\sc i} remains abundant. 
Thus  sensitive C{\sc i} 1-0, 2-1 imaging observations
 of galactic gas outflows could reveal significantly more $\rm H_2$ gas
 mass than CO lines currently find. C{\sc ii} also remains abundant
but starts strongly recombining to C{\sc i} at the high density end of
$n\sim200\,{\rm cm}^{-3}$, as expected from our previous work
\citep{Bisb17}.

In Figure~\ref{fig:panels}, we correlate the H$_2$ column density with the
T$_{\rm b}$(C{\sc i})/T$_{\rm b}$(CO) brightness ratio for the four different
ISM environments we considered. As expected, high H$_2$ column densities (i.e.
$N({\rm H}_2)\gtrsim10^{22}\,{\rm cm}^{-2}$) correspond to a small T$_{\rm
b}$(C{\sc i})/T$_{\rm b}$(CO) brightness ratio, implying that CO (1-0) is
brighter than C{\sc i} (1-0). From Figures~\ref{fig:cds} and \ref{fig:highzcr},
it can be seen that low $N$(H$_2$) corresponds to gas temperatures that may
exceed $\sim50\,{\rm K}$, particularly when the CR ionization rate is elevated
(i.e. $\zeta'\gtrsim30$). This may result in a local formation of CO via the OH
channel \citep[see][for further details]{Bisb17}, which then increases T$_{\rm
b}$(CO) and decreases T$_{\rm b}$(C{\sc i}). This in turn, lowers their
brightness ratio that we examine, resulting in a local minimum at
$N$(H$_2$)$\sim0.8-1\times10^{21}\,{\rm cm}^{-2}$ where FUV radiation is also important,
as it can be seen from panels b and c of Figure~\ref{fig:panels}.
For still higher CR energy densities (i.e. $\zeta'\sim300$), all
simulated GMCs become brighter in C{\sc i} (1-0) than in CO (1-0).

\section{Radio galaxies: Molecular gas and AGN-injected Cosmic rays}
\label{ssec:radio}

The powerful jets of radio galaxies can carry CRs to great distances
outside the galaxy where the radio-loud AGN resides, with the CRs also
diffusing around the immediate confines of the jet to form magnetized
bubbles \citep{Guo12}. It is possible that such powerful jets can also
entrain molecular gas from the ambient ISM and drive molecular gas
outflows, an effect already shown in the low-powered jets found in
AGN-harbouring spirals like NGC\,1068 \citep{Lama16}, and NGC\,4258
\citep{Krau07b}. This is not so difficult to imagine, given that such
jets are often launched from within gas-rich galaxies where the
radio-loud AGN resides. Moreover, the heavily flared star-forming $\rm
H_2$ gas disks expected around AGN \citep{Wada09} can act as a
constant source of molecular gas to be entrained by jets `firing' from
the AGN, given that the spins of disk and BH rotation will not
necessarily be aligned.

The interaction of cold gas with radio jets has already been studied
theoretically and invoked to explain the H{\sc i} outflows found in
some powerful radio galaxies \citep{Morg05,Krau07}, which have 
been since augmented by fast
molecular gas outflows observed via the traditional method of CO lines
\citep{Morg15,Morg16}. Finally, significant amounts of molecular gas, 
the fuel of
SF, driven out of radio-galaxies via jet-powered outflows provides a
natural explanation for the so-called `alignment effect' observed in
many gas-rich high-redshift radio-galaxies \citep[e.g.][]{McCa87,Pent01}.

Should molecular gas be `caught' in a radio-jet driven outflow, it will be
subjected to its withering CR-intense environment, which could quickly render
it CO-poor/invisible. Moreover, a great deal of low-density gas (i.e. the phase
where CR-induced and far-UV induced CO destruction are most effective), is to
be expected in such environments for the same reasons mentioned in
\S\ref{ssec:outflows} (to which perhaps MHD-driven shear should also be added).

\begin{figure*}
\centering
\includegraphics[width=0.8\textwidth]{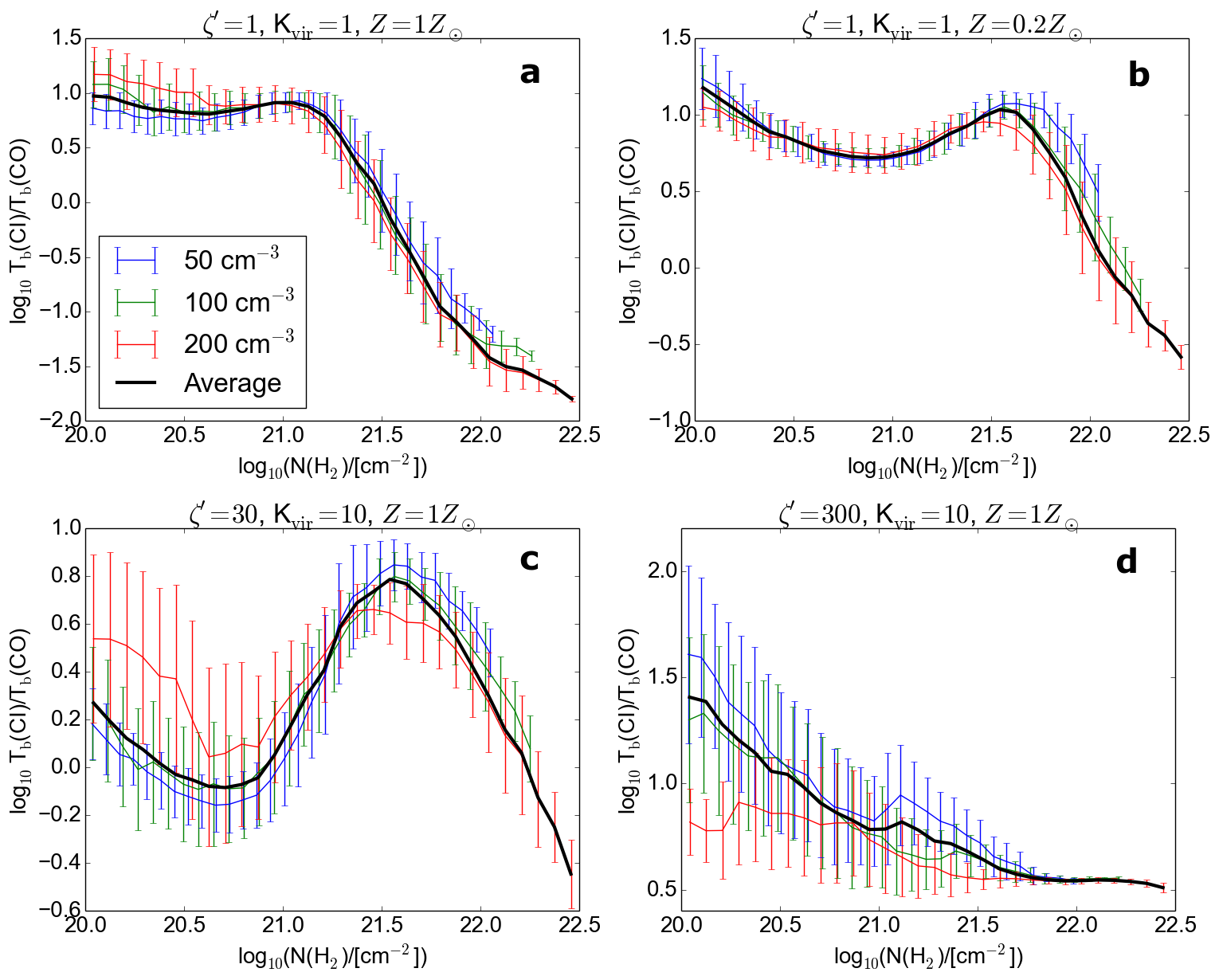}
\caption{Correlation of $N$(H$_2$) with T$_{\rm b}$(C{\sc i})/T$_{\rm b}$(CO) (for the lowest 1-0 transition) for all different ISM environments we explore. The error-bar corresponds to 1$\sigma$ standard deviation. Lines in blue color corresponds to the GMC with $\langle n\rangle\sim50\,{\rm cm}^{-3}$, green with $\sim100\,{\rm cm}^{-3}$ and red with $\sim200\,{\rm cm}^{-3}$. The black solid line shows the average value of the T$_{\rm b}$(C{\sc i})/T$_{\rm b}$(CO) ratio from all three different GMCs in each case. The local minimum in panels b and c are due to the formation of CO via the OH channel \citep{Bisb17}, since in this regime the gas temperature is $T_{\rm gas}\gtrsim50\,{\rm K}$ (see Fig.~\ref{fig:highzcr}). Note that for panels b, c and d, we always obtain T$_{\rm b}$(C{\sc i})/T$_{\rm b}$(CO)$\gtrsim0.1$. }
\label{fig:panels}
\end{figure*}

Magnetic fields in radio-jets can be strong with $\rm \langle B\rangle
\sim(35-100)\mu$G \citep{Staw05,Ostr98}. Assuming equipartition between CR and
magnetic field energy, yields CR energy densities within radio-jet environments
of $\rm U_{CR}\sim(35-280)\times\rm \langle U_{CR,Gal}\rangle$. Thus, the CR
energy density boost expected within jets and the areas near them approaches
those expected in vigorously SF galaxies, even as the source of CRs is
different. In Figure~\ref{fig:highzcr}, we show images of the relative H{\sc
i}, $\rm H_2$, CO, C{\sc i} and C{\sc ii} distributions for low-density clouds
subjected to CR-irradiation environments of $\rm U_{CR}=300\times\rm \langle
U_{CR,Gal}\rangle$, and in a strongly unbound state ($\rm K_{vir}=10$), that
are plausible for gas found in radio-jets and radio-loud AGNs. CO is very
effectively destroyed while both C{\sc i} and C{\sc ii} remain abundant for
such molecular gas (see Fig.~\ref{fig:panels}d).

Nearby radio galaxies, such as Cen~A and Minkowski's object, are
excellent targets for detecting CO-poor/C{\sc i}-rich molecular gas
in jets and their vicinity. In Cen~A, such gas may
have already been detected, even if its bright C{\sc i} emission
has been attributed to PDRs rather than CRDRs
\citep{Isra17}\footnote{{\it Relative} molecular line ratios can
 always be fitted with PDRs \citep[e.g.][]{vand10}, yet the decisive
 test whether PDRs or CRDRs/XDRs are responsible for any observed
 extragalactic lines must use the relative gas mass fractions of
 warm/dense gas and warm dust \citep[see][]{Brad03,Papa14,vand10}. 
 No such tests have been done for the C{\sc i}-bright molecular gas in Cen~A.}

In the case of Minkowski's object, sensitive CO observations detected
only a small CO-marked $\rm H_2$ gas reservoir fueling the star
formation observed along its jet, making this object an outlier of the
Schmidt-Kennicutt (S-K) relation \citep{Salo15, Lacy17}. Sensitive
[C{\sc i}] (1-0) imaging of this galaxy is particularly promising for
detecting molecular gas that may have been rendered CO-invisible by
the high $U_{\rm CR}$ expected in its radio jet. Similar observations
of radio galaxies at high redshifts are even more promising, because
in the early Universe these galaxies reside in very H$_2$-rich hosts
with plenty of gas to be entrained/impacted by their powerful jets.
Finally, [C{\sc i}] (1-0) and even (2-1) line imaging observations can
be conducted in the distant Universe with modest $T_{\rm sys}$ since
their frequencies will be redshifted to more transparent parts of the
Earth's atmosphere.

\section{Main Sequence galaxies: CRs and low-metallicity gas}
\label{ssec:ms}

A mostly CR-regulated average [CO]/[H$_2$] abundance in the ISM of Main Sequence galaxies and
the possibility of large fractions of their molecular gas mass having their CO
destroyed (with adverse effects on the calibration attempts of their X$_{\rm
CO}$-factor \citep[e.g.][]{Genz12,Carl17}) have been described by
\citet{Bisb15,Bisb17}. Our current models of lower density gas clouds,
reinforce this view by demonstrating that the conditions for a
 phase transition towards very CO-poor gas remain favorable for low densities
 {\it even for Galactic or only modestly elevated levels of CR energy densities.} For
low-metallicity gas, the effects of CO destruction are dramatic
(Figure~\ref{fig:cds}), and could explain the lack of total CO emission seen in
some metal-poor MS galaxies \citep{Genz12}. Such CO-dark molecular gas may also
exist in the outer regions of MS galaxies, which otherwise show CO-bright and
metal-rich molecular gas in their inner SF regions.

One may ask whether significant amounts of lower-density molecular gas
can indeed exist in such vigorously SF disks. Given that star
formation is a low-efficiency process, both on the mass scale of individual
molecular clouds and that of galaxy-sized reservoirs, we can
expect that large amounts of non-SF molecular gas will always be
present in SF galaxies, either in their disk, or expelled out by SF
feedback (e.g. the massive non-SF ${\rm H}_2 $ gas reservoir in the outer
regions of M~82). With CRs able to  `leak' out of the SF areas
along  molecular gas outflows and beyond the SF disk, they
can subject non-SF low density gas to significant levels of CR
irradiation, destroying its CO and replacing with C{\sc i} and C{\sc
 ii}. Sensitive C{\sc i} line imaging of such systems can show
whether this is so. Nevertheless at redshifts of $z\ga 4$, C{\sc ii}
line imaging will be the most effective in revealing such gas.

\subsection{C{\sc i} line imaging and a critical brightness limit}

There are now several PdBI and ALMA C{\sc i} (1-0), (2-1)
line observations of galaxies in the distant Universe
\citep{Walt11,Zhan14,Gull16,Popp17,Both17,Emon17} as well as local ones using
{\it Herschel} \citep{Jiao17}. In all cases, the galaxies are either
unresolved or only marginally resolved in [C{\sc i}] line emission.
Such observations do not yet allow a detailed study of relative
distributions of C{\sc i} vs CO 1-0 emission in galaxies, since
unresolved or marginally resolved emission will always be dominated by
the warm and dense star-forming $\rm H_2$ gas (where both CO and C{\sc i} are
abundant). However, \citet{Krip16} recently provided a 3$''$ resolution of 
CO(1-0) and [C{\sc i}] (1-0) for the SF galaxy NGC~253.
Intriguingly, recent high-resolution ALMA imaging of [C{\sc
 i}] (1-0) lines in a local LIRG shows significant differences between the
CO and C{\sc i}-bright $\rm H_2$ gas distribution (Zhang et al. {\it in prep.}).

High-resolution imaging of CO and [C{\sc i}] (1-0), (2-1) lines of local
LIRGS is necessary to compare their emission distribution in detail
and deduce the corresponding $\rm H_2$ mass distributions. The
resolution however {\it must be high enough to separate the  more
 compact SF molecular gas distributions from the more extended
 SF-quiescent, and perhaps lower density gas reservoirs.} The strongest requirement
for such imaging is then placed by the conditions in latter where observations yield
brightness temperature ratios T$_{\rm b}$(C{\sc i})/T$_{\rm b}$(CO) $\sim0.15$ \citep[][and references therein]{Papa04}.
From Figure~\ref{fig:panels} we can see that a C{\sc i}/CO (J=1-0) relative
brightness cutoff of T$_{\rm b}$(C{\sc i})/T$_{\rm b}$(CO)$\sim0.10$ is adequate to encompass all the $\rm
H_2$ gas in the clouds, irrespective of its thermal/chemical state and their
intrinsic average [CO]/[C{\sc i}] ratio. 
Predictably, it is the denser/colder inner regions of our low density cloud
models (and where the average $N$(H$_2$) is the highest), where this ratio
drops to its lowest values, as is indeed found by simulations and observations of
cold/low-density non-SF molecular cloud regions in the Galaxy \cite[]{Papa04,Offn14,Lo14,Glov16}.

It is perhaps more beneficial to conduct such high brightness
sensitivity C{\sc i} (and CO) line imaging for SF disks in the more
distant Universe (e.g in the disks of MS and submm galaxies, radio
galaxy environments) as to take advantage of the lower $T_{\rm sys}$
ALMA values for the C{\sc i} redshifted frequencies. Even at $z\ga
0.4$, the C{\sc i} (1-0) frequency already shifts to $\la 350$\,GHz,
where the atmosphere becomes much more transparent and detectors less
noisy. Furthermore, it is during earlier cosmic epochs (and thus
distances) when galaxies become more $\rm H_2$-rich, while the CRs
generated by their elevated SFR can induce large scale CO destruction
leaving behind C{\sc i}-rich gas.

\section{Conclusions}
\label{sec:3}

The astrochemistry that demonstrated the critical role of CRs in
regulating the average $\rm [CO]/[H_2]$ abundance for the bulk of the
molecular gas in SF galaxies (except in localised surface PDRs near
O,B stars) indicates that CO-poor/C{\sc i}-rich can exist not only in
the highly CR-irradiated ISM environments of starbursts but also in
environments with much lower levels of CR-irradiation, if the average
molecular gas density is low. A low-density molecular gas phase with
CR irradiation levels high enough to render it very CO-poor and C{\sc
 i}/C{\sc ii}-rich can be found in a number of places in the
Universe, namely:

\begin{itemize}

 \item Low-density envelopes around the CO-rich parts of ordinary
 molecular clouds in the Milky Way,

 \item Molecular gas outflows from galaxies, induced by starburst and/or AGNs,

 \item In main sequence galaxies both in their metal-rich and metal-poor regions
 (in the latter ones CR and far-UV irradiation penetrate deep in $\rm H_2$ destroying
 the CO tracer molecule).

 \item In regions inside and around radio-jets and perhaps even near the cores of
 powerful radio galaxies.

\end{itemize}

Sensitive C{\sc i} line observations of such environments can perhaps find more
molecular gas mass than the standard low-J CO line observations, provided
a relative brightness temperature ratio of T$_{\rm b}$(C{\sc i})/T$_{\rm b}$(CO)$\sim0.10$ (for $J=1-0$)
is reached in well-resolved images. GMCs in the Galaxy, distant Main Sequence and submm
galaxies, as well as areas around radio galaxies and their jets are
all excellent targets for this kind of imaging. In the later case,
Minkowski's object as well as Cygnus~A are some of the most prominent
radio-loud objects for such observations in the local Universe.

\section*{Acknowledgements}

The authors thank Ewine van Dishoeck for the useful
comments and discussion on several aspects of this work. PPP would
like to thank Ocean Divers at Sithonia, Halkidiki, Christos Douros and
Sokratis Vagiannis for providing a most useful distraction during
the last stages of this project. Z-Y.Z. and PPP acknowledge support from ERC 
in the form of the Advanced Investigator Programme, 321302, COSMICISM.

\bsp	
\label{lastpage}
\end{document}